\documentclass[preprint,english,showpacs,11pt,floatfix]{revtex4}

\usepackage{babel,amsmath,amssymb,dcolumn}
\usepackage[dvips]{graphics}

\begin{document}

\title{An alternative explanation of the conflict between \\
 $1/R$ gravity and solar system tests}
\author{Cheng-Gang Shao}
\affiliation{Department of Physics, Fudan University, Shanghai
200433, People's Republic of China }

\author{Rong-Gen Cai}
\email{cairg@.itp.ac.cn} \affiliation{Institute of Theoretical
Physics, Chinese Academy of Sciences, P.O. Box 2735, Beijing
100080, People's Republic of China}

\author{Bin Wang}
\email{wangb@fudan.edu.cn} \affiliation{Department of Physics,
Fudan University, Shanghai 200433,People's Republic of China }

\author{Ru-Keng Su}
\email{rksu@fudan.ac.cn} \affiliation{China Center of Advanced
Science and Technology (World Laboratory), P.B.Box 8730, Beijing
100080, People's Republic of China
\\Department of Physics, Fudan University, Shanghai 200433,
People's Republic of China }

\begin{abstract}
Recently the $1/R$ gravity has been proposed in order to explain
the accelerated expansion of the universe. However, it was argued
that the $1/R$ gravity conflicts with solar system tests. While
this statement is true if one views the $1/R$ gravity as an
effective theory, we find that this difficulty might be avoided if
one treats the $1/R$ term as a correction to the scalar curvature
term in the high curvature limit $R \gg \mu^2$.
\end{abstract}


\maketitle

The growing evidence indicates that the universe is undergoing a
period of accelerated expansion [1-3], which presents one of the
greatest problems in theoretical physics today: what drives the
accelerated expansion of the universe? The accelerated expansion is
usually explained through violations of strong energy condition by
introducing an extra component in the Einstein equations in the form
of dark energy with an equation of state  $\omega <-1/3$. The
simplest possibility for dark energy is a cosmological constant.
Unfortunately, this explanation is plagued with theoretical problems
such as why the observed value is 120 orders in magnitude less than
the theoretical estimates [4]. More recently there have been a
number of different attempts to modify gravity to yield accelerated
cosmologies at late times [5-10]. One of famous modifications is the
so-called 1/R gravity suggested by Carroll, Duvvuri, Trodden, and
Turner (CDTT) [5,11], with the following action
\begin{equation}
\label{eq1} S = \frac{\kappa ^2}{2}\int {d^4x} \sqrt { - g} (R -
\frac{\mu ^4}{R}) + \int {d^4x} \sqrt { - g} L_M
\end{equation}
where $\mu $ is a parameter with dimensions of mass and $L_M $ is
the Lagrangian density for matter. When the scalar curvature $R
\gg \mu ^{2}$, one expects that effect of the corrected term $\mu
^{4}/R$ can be neglected. In this case, the theory reduces to the
usual general relativity. When the corrected term $\mu ^{4}/R$ and
the Hilbert-Einstein term $R$ can be comparable to each other,
this theory significantly deviates from general relativity. In
particular, for the low curvature with $R\sim \mu ^{2}$, it was
found that the term $\mu ^{4}$/R can lead to an accelerated
expansion in late-time cosmologies [5]. If one chooses $\mu $ to
be in the order of current Hubble scale, this gravity theory can
describe current epoch of the accelerated universe very well [10].

However, it was argued that such a gravity theory is equivalent to
the Brans-Dicke theory with a vanishing Brans-Dicke parameter
$\omega $ and a potential. Base on this equivalence it was
subsequently proved by a number of authors that this theory is in
conflict with solar system tests [12-20]. Although some others
suggested the Palatini form by treating the metric and connection as
independent dynamical variables in the variational principle, which
seems to give a hope of constructing a viable model to describe
currently accelerated expansion of the universe, the post-Newtonian
approximation shows that it is still incompatible with solar system
observations [15,17,21].

For this theory to explain the cosmic acceleration, it requires
$\mu \sim H_{0}$, where $H_{0}\sim $1.5$\times $10$^{ - 33}$eV is
the current Hubble scale. In the equivalent Brans-Dick theory with
$\omega $=0, the effective mass of scalar field is in the order of
$\mu $. Since the Brans-Dicke theory has been extensively studied
in the literature and its post-Newtonian limit is well known, the
smallness of $\mu $ is obviously inconsistent with experiments
[12].

From Eq. (\ref{eq1}) it is expected that as $\mu  \to 0$, the
general relativity can be recovered. From the equivalent
Brans-Dick theory, however, the general relativity cannot be
recovered due to $\omega $=0. This seems not a consistent result.
Since we wish that in the low curvature limit (or in the cosmic
scale), the term $1/R$ makes sense; on the other hand, in the high
curvature limit (or in the solar system scale), the general
relativity can be recovered, that is the term $1/R$ can be
neglected in this case. In this sense, one may view that the term
$1/R$ in (\ref{eq1}) appears just as a corrected term to the
Hilbert-Einstein term $R$ in the high curvature limit. If we go in
this way, the post-Newtonian approximation of the theory
(\ref{eq1}) should be reconsidered since so far all investigations
about the post-Newtonian approximation on the 1/R gravity are
based on the expansion about a de Sitter space with constant
curvature, $R_{0}=3^{1 / 2} \mu ^{2}$. This is true if one views
the 1/R gravity as an effective theory and it holds for the whole
range of space-time curvature. However, we want to emphasize here
that in the high curvature limit, the term $1/R$ appears just as a
corrected term and the theory described by the action (\ref{eq1})
is not an effective theory. In this sense, $\mu ^{4 }$ can be
regarded as an expansion parameter in the high curvature limit.
Therefore the Newtonian approximation of general relativity still
holds here and the term $\mu ^{4}/R$ just gives a tiny
modification to the post-Newtonian approximation of general
relativity.

In this short note we will give the modification, due to the term
$\mu ^{4}/R$, to the post-Newtonian approximation of general
relativity. Since the effect of the term $\mu ^{4}/R$ is very tiny
in the solar system, it will not give rise to any conflict with
solar system test of gravity. This expanding technique reduced the
number of the degrees of freedom and replaced the 1/R gravity
theory (which contains an extra scalar degree of freedom
\cite{22}) with a different theory without extra degrees of
freedom. Our Lagrangian agrees to that of the $1/R$ theory only in
the first two terms, and it contains additional terms with higher
orders in $\mu^4/R$. However,  we will not include such higher
order terms in following discussion since such terms make no sense
in the high curvature limit, for the purpose of demonstration,
considering the term $1/R$ is enough.

We start by varying Eq. (\ref{eq1}) with respect to the metric,
which yields the following equations of motion
\begin{equation}
\label{eq2} (1 + \frac{\mu ^4}{R^2})R_{\mu \nu } - \frac{1}{2}(1 -
\frac{\mu ^4}{R^2})Rg_{\mu \nu } - \mu ^4(\nabla _\mu \nabla _\nu
\frac{1}{R^2} + g_{\mu \nu } \nabla ^\alpha \nabla _\alpha
\frac{1}{R^2}) = \kappa ^2T_{\mu \nu }
\end{equation}
Here the metric satisfies a system of fourth-order partial
differential equations. Eq. (\ref{eq2}) has two constant curvature
vacuum solutions. For the interest from cosmological point of view,
one usually considers the positive constant-curvature solution with
scalar curvature $R_{0}=3^{1 / 2} \mu ^{2}$, so that the universe
can accelerated expand at late times. This is the de Sitter solution
mentioned above. The fourth-order equations are very complicated and
it turns out convenient to consider the trace equation of Eq.
(\ref{eq2})
\begin{equation}
\label{eq3}
 - R + 3\mu ^4(\frac{1}{R} + \nabla ^\alpha \nabla _\alpha \frac{1}{R^2})= \kappa ^2T,
\end{equation}
where the curvature is dynamic and is equivalent to a scalar field
$\phi =1+\mu ^{4}/R^{2}$. The vacuum de Sitter solution corresponds
to the case with $\phi _{0}=4/3$.

In order to compare this theory with the solar system experiments,
one can calculate the approximately static solutions. The usual
post-Newtonian approximation is performed around the de Sitter
vacuum, where the curvature is expressed as $R=R_{0}+\delta R$.
The corresponding scalar field is expressed in the form $\phi
=\phi _{0}+\delta \phi $. $\delta R$ (or $\delta \phi $)
represents the local deviation from the background curvature
$R_{0}$ (or $\phi _{0}$) and vanishes far from the local system.
Linearizing Eq.(\ref{eq3}) requires $R_{0} \gg \delta R$. However,
this does not hold for the high curvature limit $R \gg \mu ^{2}$.
Note that the solar system just belongs to the high curvature
case. That is, the breaking down of $R_{0} \gg \delta R$ in the
solar system invalidates the post-Newtonian approximation around
the de Sitter vacuum, which is also noticed in Refs.[20,21].

Since we regard the term $\mu ^{4}/R$ as a corrected term to the
Hilbert-Einstein action $R$, we therefore expand the curvature
around the parameter $\mu $, not around the de Sitter background.
If $\mu =0$, we have from Eqs. (\ref{eq2}) and (\ref{eq3}) that
$R=-T \approx \rho $ in the lowest post-Newtonian approximation,
where we have taken $\kappa =1$ and $\rho $ the energy density of
the local system. In the local solar system, we have $\rho \gg
\mu ^{2}$ and the solution of Eq. (\ref{eq3}) can be expanded in
orders of $\mu ^{4}/ \rho ^{2}$ as
\begin{equation}
\label{eq4} R \approx \rho + \frac{\mu ^4}{\rho ^2}R_1 + \frac{\mu
^8}{\rho ^4}R_2 + \cdots.
\end{equation}
Substituting Eq. (\ref{eq4}) into Eq. (\ref{eq3}) and expanding in
orders of $\mu ^{4}/ \rho ^{2}$, we obtain the first two terms of
corrections
\begin{equation}
\label{eq5} R_1 = 3\rho + 3\rho ^2\nabla ^\alpha \nabla _\alpha
\frac{1}{\rho ^2},{\begin{array}{*{20}c}
 \hfill \\
\end{array} }R_2 = - 3R_1 - 6\rho ^4\nabla ^\alpha \nabla _\alpha \frac{R_1
}{\rho ^5}.
\end{equation}
By using the expansion Eq. (\ref{eq4}), we can rewrite the equations
of motion for metric in orders of $\mu ^{4}/ \rho ^{2}$ as
\begin{equation}
\label{eq6} R_{\mu \nu } = T_{\mu \nu } - \frac{1}{2}g_{\mu \nu } T
- \frac{\mu ^4}{\rho ^2}(T_{\mu \nu } - \frac{1}{2}g_{\mu \nu } \rho
) + \mu ^4(\nabla _\mu \nabla _\nu \frac{1}{\rho ^2} +
\frac{1}{2}g_{\mu \nu } \nabla ^\alpha \nabla _\alpha \frac{1}{\rho
^5}).
\end{equation}
Next we calculate the post-Newtonian approximation around the
Minkowskian space with $g_{\mu \nu } \approx \eta _{\mu \nu }+h_{\mu
\nu }$. To the order $\mu ^{4}/ \rho ^{2}$ and the lowest
post-Newtonian approximation, the metric satisfies the following
equations
\begin{equation}
\label{eq7}
\begin{array}{l}
 \nabla ^2h_{00} = - (1 - 3\frac{\mu ^4}{\rho ^2})\rho + \mu ^4\nabla
^2\frac{1}{\rho ^2}, \\
 \nabla ^2h_{ij} = - \delta _{ij} (1 + \frac{\mu ^4}{\rho ^2})\rho - \delta
_{ij} \mu ^4\nabla ^2\frac{1}{\rho ^2} - 2\mu ^4\partial _i \partial
_j
\frac{1}{\rho ^2}, \\
 \end{array}
\end{equation}
where the harmonic coordinate condition has been used. For
simplicity, we consider the field of a static spherically symmetric
mass source, where the mass density $\rho $ is a function of $r$
only. Then the metric can be easily integrated to give
\begin{equation}
\label{eq8}
\begin{array}{l}
 h_{00} \approx 2U + \mu ^4 / \rho ^2 \\
 h_{ij} \approx \delta _{ij} (2U - \frac{\mu ^4}{\rho ^2}) - 2\mu ^4[(\delta
_{ij} - \frac{3x_i x_i }{r^2})\frac{1}{r^3}\int_0^r {\frac{r^2}{\rho
^2}dr}
+ \frac{x_i x_i }{r^2}\frac{1}{\rho ^2}] \\
 \end{array}
\end{equation}
where we have kept the first order of $U$ and $\mu ^{4}/ \rho ^{2}$
with $U$ the Newtonian potential. It might be worth stressing here
that the expansion (\ref{eq4}) does not hold for point sources and
it should be kept in mind that the term $\mu ^{4}/ \rho ^{2 }$in
(\ref{eq8}) is always less than $U$. For the solar system, taking
the mean density as $\rho \sim 10^{ - 10}$g/cm$^{3}$, we have $\mu
^{4}/ \rho ^{2}\sim $10$^{ - 36}$, which is far smaller than the
Newtonian potential $U$. Therefore, the $1/R$ gravity is compatible
with solar system experiments in our setup.

In summary, while it was claimed that the $1/R$ gravity as an
effective theory is not consistent with the solar system
observation of gravity, we have shown that if we view the term
$1/R$ as a corrected term to the Hilbert-Einstein scalar curvature
term $R$ in the high curvature limit, namely in the case of $R \gg
\mu ^{2}$, the term $\mu ^{4}/R$ will not make any conflict with
the solar system test of gravity. We have given the modification
of the post-Newtonian approximation of the general relativity due
to the term $\mu ^{4}/R$. In our setup, the $1/R$ gravity is
compatible with the solar system test. However we would like to
stress here that these modified models of gravity contain higher
order terms of $1/R$  may or may not be viable for the expansion
of the Universe, since these higher order terms will be as large
as the $\mu ^{4}/R$ term when the Universe begins to accelerate.
Finally, it is an interesting issue to see whether in our setup
the $1/R$ gravity could be ruled out by observation of
gravitational force in the laboratory [22]. At first look from
(6), one might worry that the $1/R$  gravity will give very
strongly gravitational force at low densities since $1/\rho $
appears in (6). In fact, we can see from (4) that our
approximation holds only for the case $\mu^4/\rho^2 \ll 1$.
 When $\rho^2 \sim \mu^4$, we need to consider the complete
 effective gravity, which contains other forms of
 corrections. The main purpose of this note is just to point out
 that the usual treatment of post-Newtonian approximation of the
 $1/R $ gravity is not applicable to the solar system tests of
 gravity.

\begin{acknowledgments}

We thank the referee for his/her quite useful and helpful comments
and suggestions, which help deepen our understanding of the $1/R$
gravity. This work was partially supported by NNSF of China,
Ministry of Education of China and Shanghai Education Commission.
R.K. Su's work was partially supported by the National Basic
Research Project of China.
\end{acknowledgments}


\end{document}